\documentclass[aps,twocolumn,pra,groupedaddress,superscriptaddress,nofootinbib]{revtex4}
\usepackage{graphicx,amsmath,amssymb,amsbsy,subfigure,hyperref,bbm,times,txfonts}
\usepackage{subfigure}
\usepackage{tabularx}
\hypersetup{
   colorlinks=true,
   linkcolor=blue,
}
\graphicspath{{figure/}}
\makeatletter
\g@addto@macro\@floatboxreset\centering
\makeatother

\begin{document}

\newcommand{\ket}[1]{|#1\rangle}
\newcommand{\bra}[1]{\langle#1|}
\newcommand{\ketbra}[1]{| #1\rangle\!\langle #1 |}
\newcommand{\kebra}[2]{| #1\rangle\!\langle #2 |}
\newcommand{\id}{\mathbbm{1}}
\newcommand{\ohm}{\Omega_{\rm CQ}}
\newcommand{\rhobd}{\rho^{\vec{c}}_{AB}}
\providecommand{\tr}[1]{\text{tr}{\left[#1\right]}}
\providecommand{\sprod}[2]{\langle#1|#2\rangle}
\providecommand{\expect}[2]{\bra{#2} #1 \ket{#2}}

\title{Quantumness and speedup limit of a qubit under transition frequency modulation}

\author{Amin Rajabalinia}
\affiliation{Department of Physics, University of Guilan, P. O. Box 41335--1914, Rasht, Iran}

\author{Mahshid Khazaei Shadfar}
\affiliation{Dipartimento di Ingegneria, Universit\`{a} di Palermo, Viale delle Scienze, 90128 Palermo, Italy}
\affiliation{INRS-EMT, 1650 Boulevard Lionel-Boulet, Varennes, Qu\'{e}bec J3X 1S2, Canada}

\author{Farzam Nosrati}
\affiliation{Dipartimento di Ingegneria, Universit\`{a} di Palermo, Viale delle Scienze, 90128 Palermo, Italy}
\affiliation{INRS-EMT, 1650 Boulevard Lionel-Boulet, Varennes, Qu\'{e}bec J3X 1S2, Canada}

\author{Ali Mortezapour} \email{mortezapour@guilan.ac.ir}
\affiliation{Department of Physics, University of Guilan, P. O. Box 41335--1914, Rasht, Iran}

\author{Roberto Morandotti}
\affiliation{INRS-EMT, 1650 Boulevard Lionel-Boulet, Varennes, Qu\'{e}bec J3X 1S2, Canada}

\author{Rosario Lo Franco}  \email{rosario.lofranco@unipa.it}
\affiliation{Dipartimento di Ingegneria, Universit\`{a} di Palermo, Viale delle Scienze, 90128 Palermo, Italy}

\begin{abstract} 

Controlling and maintaining quantum properties of an open quantum system along its evolution is essential for both fundamental and technological aims. We assess the capability of a frequency-modulated qubit embedded in a leaky cavity to exhibit enhancement of its dynamical quantum features. The qubit transition frequency is sinusoidally modulated by an external driving field. We show that a properly optimized quantum witness effectively identifies quantum coherence protection due to frequency modulation while a standard quantum witness fails. We also find an evolution speedup of the qubit through proper manipulation of the modulation parameters of the driving field. Importantly, by introducing a new figure of merit $\mathfrak{R_g}$, we discover that the relation between Quantum Speed Limit Time (QSLT) and non-Markovianity depends on the system initial state, which generalizes previous connections between these two dynamical features. 
The frequency-modulated qubit model thus manifests insightful dynamical properties with potential utilization against decoherence.

\end{abstract}

\date{\today }

\maketitle

\section{Introduction}

The coherent superposition of quantum states is termed quantum coherence and is regarded as the main distinguishing feature of the quantum world which has no classical counterpart. Many quantum-enhanced technologies are founded on this feature \cite{colloquium, 13.Walls2008,SunPNAS}. Tomographic methods are usually employed to detect quantum coherence by reconstructing the density matrix of quantum systems. Albeit the implementation of such a strategy poses a technical challenge as to the measurement settings of the experiment \cite{nielsen2002quantum,d2003quantum}. To overcome the complexity of detection in the experiment, Leggett-Garg inequality \cite{leggett1985quantum} and quantum witness \cite{18.Lambert2012,19.Kofler} have been introduced as quantum indicators to quantify the nonclassicality of a system. Quantum witness is based on the classical no-signaling-in-time assumption, according to which a prior experiment does not disturb the statistical outcome of the subsequent experiment \cite{18.Lambert2012,19.Kofler,clemente2016no}. Hence, in recent years, finding a standard quantum witness has been extensively studied in a wide variety of physical systems \cite{20.Schild2015,39.Wang2017,17.Friedenberger2017,knee2018subtleties,ban2019leggett,bojer2019quantum,14.Gholipour2020,21.Farzam2020,ku2020experimental,usui2020temporal,gong2020quantum,kurashvili2021quantum,karthik2021leggett}. Nonetheless, newly a pitfall is emerged in the standard quantum witness which demonstrates this indicator fails to detect the maximal violations of classicality of a moving qubit in a dissipative cavity. Therefore, in order to remove the shortcoming, the optimized quantum witness \cite{21.Farzam2020} with the suitable intermediate blind measurement has been put forward which can reach the maximum violation of its amplitude, coinciding with the coherence monotone.

Since a realistic quantum system inevitably interacts with the environment, such interactions consequently give rise to coherence loss. Usually, the bath is considered to play a destructive role in quantum coherence, with the well-known phenomenon of decoherence \cite{breuer2002theory}. As a result of decoherence, time constraint is imposed on quantum tasks. Therefore, various strategies have been devised to control and maintain quantum resources (such as coherence and entanglement) including dynamical decoupling \cite{Viola1998, viola2005random, franco2014preserving, DARRIGO2014211, orieux2015experimental, damodarakurup2009experimental, cuevas2017cut,FacchiPRA}, decoherence-free subspaces \cite{zanardi1997noiseless, lidar1998decoherence}, error correction \cite{preskill1998reliable, knill2005quantum, plenioPRA1997,shor1995scheme,LaflammePRL}, topological properties \cite{kitaev2003fault, freedman2003topological}, structured quantum and classical environments \cite{mazzola2009sudden,lofrancoreview,lofrancoQIP,aaronsonPRA,lofrancoPRA2012,xuNatComm, bellomo2008entanglementtr,BellomoEPL,Mortezapour_2017,ManOptExp,ScalaPRA,Man2015a,ManPRA2015,silvaPRL,27.Daniel2017,carusoRMP,28.Aolita2015,TanPRA,TongPRA,ScalaEPJD}, and indistinguishability-based protection \cite{nosrati2020robust, nosrati2020dynamics, piccolini2021entanglement,Perez-Leija2018,piccoliniOSID}. In fact, optimizing such quantum techniques paves the way for the realization of quantum information applications.

The minimal time during which a quantum system evolves between two extinguishable states is recognized as the quantum speed limit time (QSLT). This interval determines the maximal rate of evolution that a quantum system can reach. Regarding the literature, Leonid Mandelstam and Igor Tammthe had a pioneering role in formulating the QSLT concept by means of time-energy uncertainty relation which bounded the speed of evolution reads as MT bound \cite{mandelstam1991uncertainty}. Later, Norman Margolus and Lev Levitin refined this relation and derived a more rigorous relation (ML bound) according to which the speed of evolution cannot exceed the mean energy \cite{margolus1998maximum,deffner2013quantum}. The MT and ML bounds had been applicable for closed quantum systems, despite the fact that the realistic quantum systems are not isolated. It was of crucial importance to generalize the notion of QSLT to open quantum systems. Hence, Deffner and Lutz \cite{deffner2013quantum} filled this gap by introducing a unified quantum speed limit time. On the other hand, since in the open quantum systems the environmental effects cannot be ignored, this question is posed that how non-Markovianity can affect the QSLT. Although the early studies suggested that non-Markovianity can speed up the quantum evolution \cite{deffner2013quantum}, the recent reports indicate the existence of a direct relationship between the QSLT and memory effect is still open to debate for the most general dynamics of open quantum systems. Thus, increasing the non-Markovianity, quantified by the backflow of information \cite{breuer2009measure}, does not necessarily lead to the speedup in the evolution \cite{teittinen2019there}. 
This minimum bound also determines the maximum rate of quantum information \cite{bekenstein1981energy}, computation \cite{lloyd2000ultimate}, entropy production \cite{deffner2010generalized}, the ultimate precision in quantum sensing \cite{giovannetti2011advances}, and scrambling the spectral form factor \cite{del2017scrambling}. Moreover, the QSLT serves as the inherent limitations of quantum optimal control algorithms \cite{caneva2009optimal}.

Recent studies show that modulating the transition frequency of a qubit provides the foundation for the observation of a wide variety of novel phenomena, such as the formation of sidebands transitions \cite{Strand220505, beaudoin2012first} the coherent destruction of tunneling \cite{grossmann1991coherent}, dynamic Stark effect \cite{alsing1992dynamic}, Landau–Zener–Stückelberg-interference \cite{shevchenko2010landau}, modifying fluorescence spectrum \cite{ficek2001fluorescence}, and the enhancement of the degree of non-Markovianity \cite{poggi2017driving}. Moreover, this scheme allows observing topological transitions \cite{martin2017topological} and population trapping \cite{gray1978coherent}. On the other hand, one can employ the frequency modulation method to decouple a qubit system from its environment \cite{viola1999dynamical, agarwal1999control,kofman2001universal} and consequently protect quantum resources from unwanted environmental noise \cite{mortezapour2018protecting, nourmandipour2021entanglement}. In addition, frequency modulation of the qubit can be harnessed to create fast two-qubit gates for quantum information processing \cite{beaudoin2012first, chu2020realization}. It is noteworthy that exertion of an external off-resonant field can be employed as a tactic to perform a frequency modulation in an atomic qubit \cite{54.Griffith1998, 56.Silveri2017}. Moreover, regarding the most recent reports, the realization of the frequency modulation in superconducting qubits is feasible with the aid of state-of-the-art experimental techniques \cite{59.Tuorila2010, 57.Nakamura2001, 60.Tuorila2013, oliver2005mach, 61.Quantumcomputing}.

 In the present work, we consider a frequency-modulated qubit inside a leaky cavity for pursuing two goals: (i) checking the consistency between the standard quantum witness (SQW) and coherence monotone, motivated by the results introduced in Ref.~\cite{21.Farzam2020}; (ii) studying the role of frequency modulation in the speedup evolution, addressing the relationship between QSLT and non-Markovianity as a highly controversial issue which depends on the different initial states. Concerning the first goal, the results suggest that SQW largely fails to indicate the non-classicality of the system, certifying previous results \cite{21.Farzam2020}. Moreover, it is revealed that SQW either lags behind or exceeds the coherence monotone. Differently, the optimized quantum witness (OQW) \cite{21.Farzam2020} exhibits maximum values perfectly coinciding with the coherence monotone for optimal parameters of modulation (frequency and amplitude): this feature stresses the requirement of optimization to get a reliable quantum indicator. Regarding the second goal, we demonstrate that the system can reach a high evolution speedup for optimal modulation parameters. The achieved results indicate that the former conception of the relation between QSLT and non-Markovianity needs to be amended by introducing a new parameter, that we name $\mathfrak{R_g}$, which is interestingly related to the ratio between non-Markovianity and the population of the excited state. 

The paper is organized as follows: In Sec.~\ref{sectwo}, we present the Hamiltonian of a single frequency-modulated qubit and discuss the state evolution of the system. In Sec.~\ref{secthree}, we give a basis for assessing the quantum witness and its optimization. This section is followed by providing results that demonstrate the efficiency of OQW in measuring the quantumness of the frequency-modulated qubit. Sec.~\ref{secfour} is devoted to the discussion about the QSLT of the system and its connection to non-Markovianity. Finally, we present an outline of the main conclusions and prospects in Sec.~\ref{secfive}.

\section{Model and System}

\label{sectwo}
\begin{figure}[t!]
\begin{center}
\includegraphics[width=0.45\textwidth]{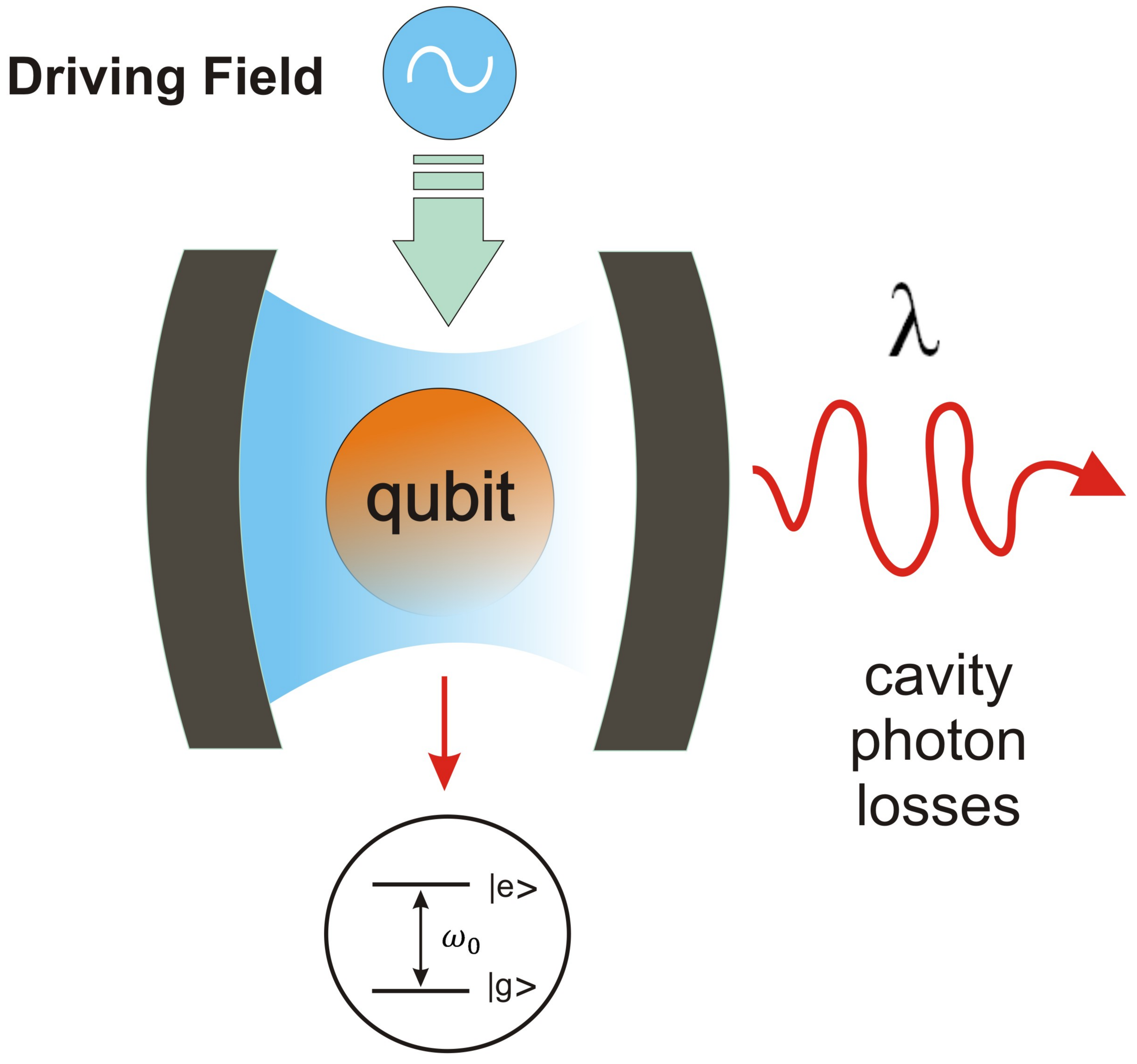}
\end{center}
\caption{Sketch of the single driven qubit system. A qubit (two-level atom) is embedded inside a structured leaky cavity. The qubit transition frequency $\omega_0$ is sinusoidally modulated via an external applied field with a modulation amplitude $\delta$ and a modulation frequency $\Omega$. The qubit interacts with vacuum modes.} 
\label{AR1}
\end{figure}

The system under consideration is a qubit (two-level system) coupled to a zero-temperature reservoir which is comprised of the quantized modes of a high-$Q$ cavity. It is assumed that the transition frequency of the qubit is modulated sinusoidally by an external driving field, as depicted in Fig.~\ref{AR1}. By utilizing the electric-dipole and rotating-wave approximations, the Hamiltonian of the system can be written as ($ \hslash=1 $):
\begin{eqnarray}\label{eq:1}
\hat{H}=\sum_{k}\omega_{k}\hat{b}_{k}^{\dagger} \hat{b}_{k} +\frac{1}{2}[\omega_{0}+\delta \cos (\Omega t)]\hat{\sigma }_{z} \nonumber \\
+ \sum_{k}\{g_{k}\hat{\sigma}_{+}\hat{b}_{k}+g_{k}^{*}\hat{\sigma}_{-}\hat{b}_{k}^{\dagger}\}.
\end{eqnarray}
where $\hat{\sigma}_{z}=\left|e\right\rangle\left\langle{e}\right|-\left|g\right\rangle\left\langle{g}\right|$ is a Pauli operator for the qubit, $\hat{\sigma}_{\pm}$ represent the raising and lowering operators of the qubit, $\hat{b}_{k}^{\dagger}$ and $\hat{b}_{k}$ are respectively the creation and the annihilation operators for a photon of mode $k$ with the frequency $\omega_{k}$ and $g_{k}$ being the coupling strength for this mode to the qubit. Furthermore, $\omega_{0}$ characterizes the transition frequency of the qubit (see Fig.~\ref{AR1}) in the absence of modulation, while $\delta$ denotes the modulation amplitude with frequency $\Omega$. After switching to a non-uniformly rotating frame (interaction picture) by the unitary transformation
\begin{equation}\small \label{eq:2}
\hat{U}=\exp \left[-i\left\{\sum _{k}\omega _{k} \hat{b}_{k}^{\dagger} \hat{b}_{k}t + [\omega _{0}t +(\delta /\Omega )\sin (\Omega t)]\hat{\sigma }_{z} /2\right\}\right]. 
\end{equation}
One can obtain the effective hamiltonian $\hat{H}_\mathrm{eff} =\hat{U}^{\dagger} \hat{H}\hat{U}+i(\partial \hat{U}^{\dagger} /\partial t)\hat{U}$ as
\begin{eqnarray}\label{eq:3}
\hat{H}_\mathrm{eff} &=& \sum _{k}g_{k} \hat{\sigma }_{+} \hat{b}_{k} e^{-i(\omega _{k} -\omega _{0} )t} e^{i(\delta /\Omega )\sin (\Omega t)} \nonumber \\
&+& \sum _{k} g_{k}^{*} \hat{b}_{k}^{ \dagger } \hat{\sigma }_{-} e^{i(\omega _{k} -\omega _{0} )t} e^{-i(\delta /\Omega )\sin (\Omega t)}\ .
\end{eqnarray} 
Note that the exponential factors on the right-hand side of Eq.~(\ref{eq:3}) are periodic therefore, they can be rewritten as below with making use of the Jacobi-Anger expansion.
\begin{equation}\label{eq:4}
e^{\pm i(\delta /\Omega )\sin (\Omega t)} =J_{0} \left(\frac{\delta }{\Omega } \right)+2\sum _{n=1}^{\infty }(\pm i)^{n} J_{n} \left(\frac{\delta }{\Omega } \right) \cos (n\Omega t), 
\end{equation} 
where $J_{n} \left(\frac{\delta }{\Omega }\right)$ is the Bessel function of the first kind with order $n$. Starting from the initial state $\left| \Psi (0) \right\rangle ={\cos(\frac{\theta}{2}) \left| e \right\rangle +\sin(\frac{\theta}{2})e^{i\phi} \left| g \right\rangle} \left| 0 \right\rangle$, where $\left| 0 \right\rangle$ is the vacuum state of the cavity, the conservation of the number of excitations gives the following time evolution of the whole system:
\begin{eqnarray}\label{eq:5}
{\left| \Psi (t) \right\rangle} =\cos(\theta/2) C(t) {\left| e \right\rangle} {\left| 0 \right\rangle} +\sin(\theta/2)e^{i\phi} {\left| g \right\rangle} {\left| 0 \right\rangle}\nonumber \\
+{\sum _{k}C_{k} (t){\left| g \right\rangle} {\left| 1_{k}  \right\rangle}}, 
\end{eqnarray} 
where ${\left| 1_{k}  \right\rangle}$ describes a single photon in mode $k$ of the reservoir and $C_{k} (t)$ represents its probability amplitude. Solving the Schrödinger equation for the total system results in the following differential equations for the probability amplitudes $C(t)$ and $C_{k} (t)$   
\begin{equation}\label{eq:6}
\dot{C}(t)=-i (i/\alpha)\exp[i(\delta /\Omega)\sin (\Omega t)]\sum _{k}g_{k} e^{-i(\omega _{k} -\omega _{0} )t} C_{k} (t) ,                            
\end{equation}
\begin{equation}\label{eq:7}
\dot{C}_{k} (t)=-i\alpha \exp[-i(\delta /\Omega )\sin (\Omega t)] g_{k}^{*} e^{i(\omega _{k} -\omega _{0} )t} C(t).
\end{equation} 
Successively solving Eq.~\ref{eq:7} formally and substituting the achieved result into Eq.~\ref{eq:6}, yields
\begin{equation}\label{eq:8} 
\dot{C}(t)+\int _{0}^{t}dt'  F(t,t' )C(t')=0, 
\end{equation}
where the kernel $F(t,t' )$ is the correlation function defined in terms of continuous limits of the environment frequency 
\begin{eqnarray}\label{eq:9}
F(t,t' )=\exp[i(\delta /\Omega)\{\sin(\Omega t) -\sin(\Omega t')\}] \nonumber \\
\times\int _{0}^{\infty}J(\omega_{k})e^{-i(\omega_{k}-\omega_{0})(t-t')}d\omega_{k}. 
\end{eqnarray}
Here $J(\omega_{k})$ indicates the spectral density of reservoir modes. Since we consider the case in which the qubit resonantly interacts with the cavity modes, possessing Lorentzian spectral distribution, $J(\omega_{k})$ adopts the following form \cite{30.breuer2002theory,Lofranco2013}:
\begin{equation}\label{eq:10} 
J(\omega)=\frac{1}{2\pi } \frac{\gamma \lambda ^{2} }{[(\omega_{0} -\omega_{k})^{2} +\lambda ^{2}]}, 
\end{equation} 
with $\lambda$ the decay rate and $\gamma $ the spectral width of the coupling \cite{30.breuer2002theory}. Making use of the Lorentzian spectral density, the kernel $F(t,t' )$ can be acquired as follows:
\begin{equation}\label{eq:11} 
F(t,t' )=\frac{\gamma \lambda }{2} e^{-\lambda (t-t')} 
e^{[i(\delta /\Omega )\{ \sin (\Omega t) -\sin (\Omega t')\}]}. 
\end{equation}
By substituting it into Eq.~(\ref{eq:7}), one gets
\begin{eqnarray}\label{eq:12}
\dot{C}(t)+\frac{\gamma \lambda}{2} e^{[i(\delta/\Omega )\sin (\Omega t)]}\hspace{10em}\nonumber \\                                                                   
\times\int _{0}^{t}dt' e^{[-i(\delta /\Omega )\sin (\Omega t')]}e^{-\lambda (t-t' )} C(t')=0.\hspace{2em}
\end{eqnarray} 
the probability amplitude $C(t)$ can be obtained by numerically solving the above equation. Then, the reduced density matrix of the qubit in the atomic basis $\{\left| e \right\rangle, \left| g \right\rangle\}$ can be expressed as            
\begin{equation}\label{eq:13}
\rho^{q} (t)=
\left(
\begin{array}{cc}
{\cos^2(\frac{\theta}{2}) \left|C(t)\right|^{2} } & {\frac{1}{2}\sin(\theta)e^{-i\phi} C(t)} \\
{\frac{1}{2}\sin(\theta)e^{i\phi} C^{*} (t)} & {1-\cos^2(\frac{\theta}{2}) \left|C(t)\right|^{2} } \\
\end{array}
\right).
\end{equation}

With the evolved reduced density matrix above we can carry on our analysis, as reported in the following sections.

\section{Standard and Optimized quantum witness }\label{secthree}
In this section, we apply the quantum witness as a criterion of nonclassicality to quantify the quantumness of a frequency-modulated qubit inside a leaky cavity. Our general aim is to demonstrate that SQW is no longer a reliable quantum indicator for non-isolated systems and needs to be refined as the Optimized Quantum Witness (OQW). Hence, the proposed scheme best exemplifies the pitfall associated with SQW and highlights the importance of such refinement one more time as it has already been reported in ref \cite{21.Farzam2020}. Let us mention that the attributed calculations are performed considering the no-signaling-in-time condition and the results suggest that SQW fails to detect the non-classicality of a frequency-modulated qubit in a dissipative cavity.

\subsection{Standard Quantum Witness}
In the present subsection, we intend to analyze the nonclassicality of the frequency-modulated qubit by employing the SQW which is defined as 
\begin{equation}\label{eq:14}
W_{q}(t)=\left|P_{m}(t)-P'_{m}(t)\right|.
\end{equation}
where $P_{m}(t)$ denotes the quantum probability of finding the system in the state $m$ at time $t$ without performing any prior measurement, while $P'_{m}(t)=\sum _{n=1}^{d }P(m,t|n,t_{0})P_{n}(t_{0})$ represents the classical probability of finding the system in state $m$ at time $t$ before which an intermediate nonselective measurement of the state $n$ has been performed at time $t_{0}$. According to the classical no-signaling-in-time assumption, the prior measurement at time $t_{0}$ is noninvasive and does not perturb the statistical outcome of the subsequent measurement at time $t$, thus $P_{m}(t)=P'_{m}(t)$, $[W_{q}=0]$ and the system acts as a classical one. However, inequality of these two probabilities namely nonzero values of $W(t)$ violate this assumption and signify the nonclassicality of the system state at time $t$. Also, the upper bound of the standard quantum witness \cite{20.Schild2015} is $W_{q}^{max}(\tau)=1-1/d$ where $d$ is the system dimension and describes the number of possible outcomes of a blind measurement. Before proceeding to calculate the quantum witness, it is convenient to initially find a propagator for the reduced density matrix of the qubit owing to the fact that the quantum and classical probabilities are acquired by averaging projection operators on the system state at time $t$. One can obtain the propagator $\aleph$ using Lindblad-type evolution for an operator $\hat{X}$ in the Heisenberg picture $d\hat{X}/dt=L[\hat{X}]$ \cite{30.breuer2002theory,17.Friedenberger2017}. For a dissipative system-environment model, the integro-differential equation reads as 
\begin{equation}\label{eq:15}
\hat{X}(t)+\int _{0}^{t}dt'K_{t}[\hat{X}(t')]=0,
\end{equation}
where
\begin{equation}\label{eq:16}
    \begin{split}
      K_{t}[X(t')]&=F(t, t')(\sigma_{+}\sigma_{-}X(t')+X(t')\sigma_{+}\sigma_{-}\\ 
&-2\sigma_{+}\hat{X}(t')\sigma_{-}).
\end{split}
\end{equation}
Let us recall that the function $F(t, t')$ has already defined in Eq.~(\ref{eq:11}). Considering the evolution of the basis of Pauli operators $\{I,\sigma_{x},\sigma_{y},\sigma_{z}\}$,
\begin{equation}\label{eq:17}
\left(
\begin{array}{cc}
{\sigma_{x}(t)} \\
{\sigma_{y}(t)} \\
{\sigma_{z}(t)} \\
{I(t)}  \\
\end{array}
\right)=\aleph(t,0)\left(
\begin{array}{cc}
{\sigma_{x}(0)} \\
{\sigma_{y}(0)} \\
{\sigma_{z}(0)} \\
{I(0)}\end{array}
\right),
\end{equation}
where
\begin{eqnarray} \label{eq:18} 
&\aleph(t,0)=\hspace{12em}&\nonumber \\                                                                   
& \small \left(
\begin{array}{cccc}
{\frac{1}{2}(C(t)+C^{*}(t))} & {\frac{-i}{2}(C(t)-C^{*}(t))} & {0} & {0} \\
{\frac{1}{2}(C(t)-C^{*}(t))} & {\frac{i}{2}(C(t)+C^{*}(t))} & {0} & {0} \\
{0} & {0} & {{\left|C(t)\right|}^2} & {{\left|C(t)\right|}^{2}-1} \\
{0} & {0} & {0} & {1}\\
\end{array}
\right).&
\end{eqnarray}
Then, finding the average values of the Pauli operators at a time $t$, such that $\left\langle\sigma_{i}(t)\right\rangle=\aleph(t, 0)\left\langle\sigma_{i}(0)\right\rangle, (i=x, y, z)$, the qubit density matrix at time $t$ is obtained as
\begin{equation}\label{eq:19}
P(t)=\frac{1}{2}{\left(I+\left\langle\sigma_{x}(t)\right\rangle\sigma_{x}+\left\langle\sigma_{y}(t)\right\rangle\sigma_{y}+\left\langle\sigma_{z}(t)\right\rangle\sigma_{z}\right)}.
\end{equation}
The quantum probability $P_{\pm}$ of finding the state $\left|{\pm}\right\rangle$ at time $\tau$ in the absence of the intermediate nonselective measurement is given by
\begin{equation}\label{eq:20}
P_{\pm}(\tau)={\left\langle\Pi_{\pm}(\tau)\right\rangle}=Tr\left(P(\tau)\Pi_{\pm}(\tau)\right).
\end{equation}
where $P(\tau)$ is the evolved reduced density matrix of the qubit. We let ${\Pi_{\pm}^{x}}=\left(I\pm\sigma_{x}\right)/2$ \cite{17.Friedenberger2017}, be the intermediate nonselective projections operators and assume the qubit is initially in a coherent superposition $\left| \Psi (0) \right\rangle ={\cos(\frac{\theta}{2}) \left| e \right\rangle +\sin(\frac{\theta}{2})e^{i\phi} \left| g \right\rangle} \left| 0 \right\rangle$. Hence, the quantum and classical probabilities $P_{\pm}(\tau), P'_{\pm}(\tau)$ can be used to calculate the SQW according to 
\begin{eqnarray}\label{eq:21}
&W_{q}(\tau)=\left|P_{+}(\tau)-P'_{+}(\tau)\right| & \nonumber \\
&=\frac{1}{4}{\left|\sin(\theta)\left(C(\tau)+C^{*}(\tau)-\frac{1}{2}{\left(C(\tau/2)+C^{*}(\tau/2)\right)}^{2}\right)\right|}.&
\end{eqnarray}

\begin{figure}[t!]
\begin{center}
\includegraphics[width=0.5\textwidth]{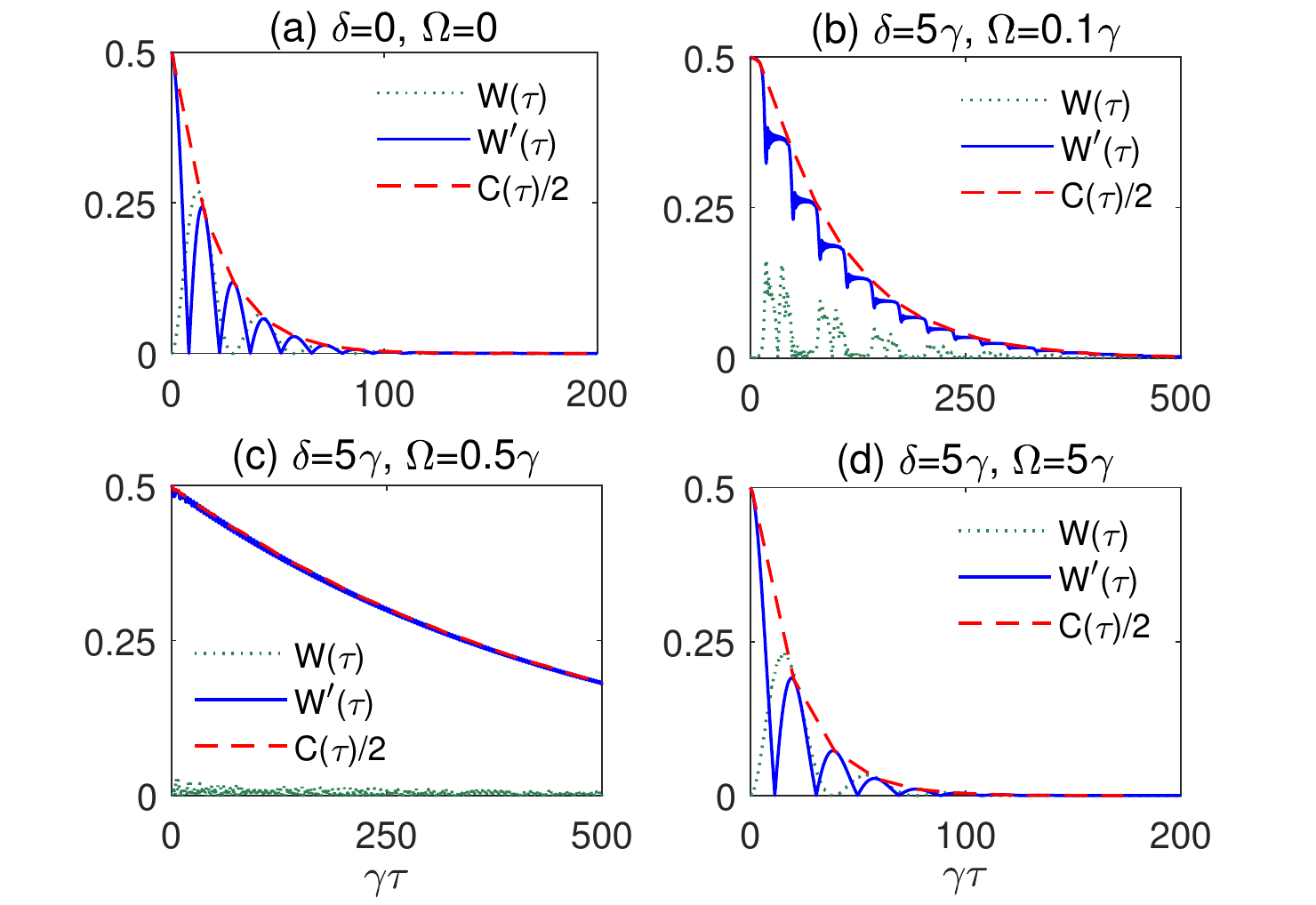}
\end{center}
\caption{Standard Quantum witness $W(\tau),$ Optimized quantum witness $W'(\tau),$ and Coherence monotone $C(\tau)/2,$ as a function of the dimensionless time interval $\gamma\tau$ for different values of modulation frequency $\Omega$ and a fixed modulation amplitude $\delta=5~\gamma$. The panel $(a)$ corresponds to the unmodulated condition $\delta=0$, $\Omega=0$, $(b)$ $\Omega=0.1\gamma$, $(c)$ $\Omega=0.5\gamma$ and $(d)$ $\Omega=5\gamma$. Other parameters are $\lambda=0.1\gamma$, $\theta=\pi/2$ and $\phi=0$.}
\label{AR2}
\end{figure}

\begin{figure}[t!]
\begin{center}
\includegraphics[width=0.5\textwidth]{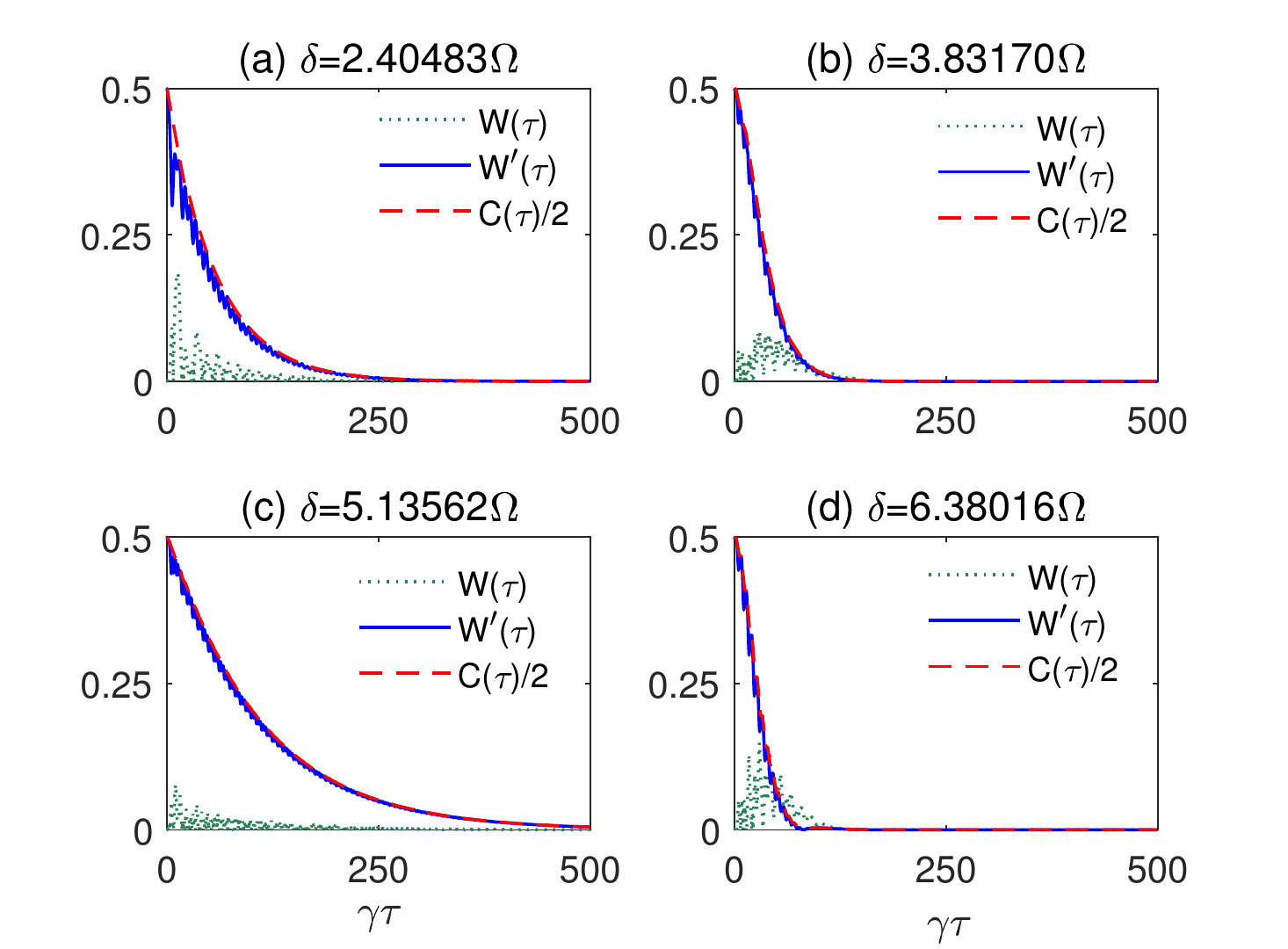}
\end{center}
\caption{Standard Quantum witness $W(\tau),$ Optimized quantum witness $W'(\tau),$ and Coherence monotone $C(\tau)/2,$ as a function of the dimensionless time interval $\gamma\tau$ with various values of the modulation frequency and amplitude: $(a)$ $\delta=2.40483\Omega~[J_0 (2.40483)=0]$, $(b)$ $\delta=3.83170\Omega~[J_1 (3.83170)=0]$, $(c)$ $\delta=5.13562\Omega~[J_2 (5.13562)=0]$, $(d)$ $\delta=6.38016\Omega~[J_3 (6.38016=0]$. Other parameters are $\lambda=0.1\gamma$, $\theta=\pi/2$ and $\phi=0$.}
\label{AR3}
\end{figure}

\subsection{Optimized Quantum Witness}
It has been demonstrated that the SQW of an isolated system can reach its maximum value equivalent to the coherence monotone $W_{q}^{max}(\tau)={C(\tau)}/2$ in which ${C(\tau)}$ is the envelope of quantum coherence. One can simply obtain ${C(\tau)}$ by employing the $\ell_{1}-$ norm of coherence $C_{\ell_{1}}(\rho)=\sum_{i\ne{j}}|\left<i|\rho|j\right>|$ \cite{63.Quantifyingcoherence}. The same maximal value of quantum witness has been also validated for a damped qubit in a Markovian thermal reservoir. Namely the envelope of the quantum witness, defined by means of the usual intermediate and final measurements $\Pi_{\pm}^{x}$ of Eq.~(\ref{eq:20}), perfectly matches the coherence monotone \cite{20.Schild2015}. However, the most recent study concerning quantum witness poses a challenging question whether for a generic nonisolated quantum system in which the non-Markovian behavior appears it is guaranteed that quantum witness coincides with the coherence monotone. Their results suggest that intermediate blind measurement matters in determining the upper bound of SQW and they optimize the quantum witness by substituting $\Pi_{\pm}^{z}=\frac{1}{2}{\left(I\pm\sigma_{z}\right)}$ for the conventional blind measurement i.e., $\Pi_{\pm}^{x}=\frac{1}{2}{\left(I\pm\sigma_{x}\right)}$ \cite{21.Farzam2020}. The proposed system once again testifies that such optimization is crucially required. In this context, for the proposed dissipative system following the $\ell_{1}-$ norm of coherence we simply obtain $C_{\ell_{1}}=\left|\sin(\theta)C(t)\right|$ from the qubit reduced density matrix of Eq.~(\ref{eq:13}). Moreover, we apply the operator ${\Gamma(\rho(\tau/2))}=\sum_{i}{\left|i\right>\left<i\right|\rho(\tau/2)\left|i\right>\left<i\right|}$, to classicalize the system which preserves the diagonal entries of the system state and discard the off-diagonal ones \cite{62.Knee2018}. 
One can rewrite the Eq.~(\ref{eq:20}) by employing the optimal nonselective projections. The operations effectively make the perturbed intermediate state classical, so that any incoherent channel remains classical for the rest of the evolution. It is analogous to maximizing the distance between the state of the system at $t=\tau$ and its perturbed counterpart. Calculating these new blind measurements and having the perturbed state ${\rho'(\tau/2)}$ for the subsequent evolution at the time leads to the qubit state at ${t=\tau}$
\begin{equation}\label{eq:22}
{\rho'(\tau)}=\left(
\begin{array}{cc}
{\cos^2{(\frac{\theta}{2})}{\left|C(\tau)\right|}^4} & {0} \\
{0} & {1-\cos^2{(\frac{\theta}{2})}{\left|C(\tau)\right|}^4} \\
\end{array}
\right), \\  
\end{equation} 
where ${\rho'(\tau)}$ is the reduced density matrix of the qubit after measuring the probability amplitude ${C(\tau)}$. Accordingly, the quantum and classical probabilities are obtained with the usual final measurement set by projector $\Pi_{x,+}$. The optimized quantum witness of the system under a well-defined blind measurement is 
\begin{equation}\label{eq:23}
W'(\tau)=\frac{1}{4}{\left|\sin(\theta)(C(\tau))+C^{*}(\tau))\right|=\frac{1}{2}\left|\sin(\theta)Re[C(\tau)]\right|}.
\end{equation}
The optimization procedure on the standard quantum witness ${W_{q}(\tau)}$ of Eq.~(\ref{eq:21}) coincides with the real part of the off-diagonal term of the evolved qubit density matrix. In order to draw a comparison among the dynamic behavior of the SQW (${W_{q}(\tau)}$), the OQW (${W'_{q}(\tau)}$) and the coherence monotone with respect to the amplitude and frequency modulation parameters (respectively, $\delta$ and $\Omega$), we plot them simultaneously in Fig.~\ref{AR2} as a function of the dimensionless time interval $\gamma\tau$. The time evolution of the system is compared to the case where the external driving is off ${(\delta=0,\Omega=0)}$ (see Fig.~\ref{AR2}(a)), for which coherence monotone monotonically decreases, and both ${W_{q}(\tau)}$ and ${W'_{q}(\tau)}$ manifest this behavior well. Such behavior is also observed in panel (d) when the modulation frequency is high. However, panels (b) and (c) reveal the inconsistency between the pattern of the behavior of SQW and the coherence monotone. So that, the SQW curve is unable to touch the coherence monotone however, the OQW curve keeps up with the coherence monotone. Such inconsistency is abundantly clear in panel (c), for the optimal value of the frequency modulation, in which SQW approximately possesses the constant zero value whereas OQW perfectly matches the coherence monotone.

In order to gain a far better understanding of the dynamic of the system, we subsequently aim to compare SQW and OQW when the ratio of amplitude and frequency of modulation parameter $(\delta/\Omega)$ is intentionally tuned to the first zero of the n-th Bessel function ${J_{n}}$ in Fig.~\ref{AR3}. By taking a sweeping glance at all the panels, one would notice that SQW either lags behind or exceeds (when the ratio $\delta/\Omega$ is adjusted to the first zero of the third Bessel function) the coherence monotone. Whereas, the OQW amplitude reaches its upper limit, being exactly equal to the coherence monotone. Regarding such glaring inconsistency between the behavior of SQW and coherence monotone, one can easily deduce that the SQW is truly incapable of detecting the behavior of the generic quantum systems in which non-Markovian environmental effect appears. Simultaneously, it is proved that the adaptive blind measurement $\Pi_{\pm}^{z}$ is a faithful measure to classify quantum from classical behavior in the experiment. Moreover, the proposed model not only certifies the violation of SQW \cite{62.Knee2018} but, in contrast to the previous reports in which SQW could not reach the coherence monotone, it also reveals the SQW can exceed the coherence monotone. The disclosure of such serious shortcomings in SQW stresses the crucial requirement of optimization.

\section{Interplay between QSLT and non-Markovianity of a frequency-modulated qubit}\label{secfour}
The primary purpose of this section is to assess the impact of frequency modulation on the speedup evolution of a qubit embedded in a leaky cavity. Meanwhile, two questions immediately arise: (a) What is the connection between QSLT and non-Markovianity in the modulated quantum system? (b) Would non-Markovianity be the sole factor in speeding up the rate of evolution of the frequency-modulated system or do other contributing factors participate? To address these questions, quantum speed limit time (QSLT) can be employed to analyze the maximal speed of the evolution of the frequency-modulated qubit. Initially, it is required to provide preliminary explanations concerning QSLT. The QSLT determines the minimal time required by the system to completely evolve between two distinguishable states \cite{mandelstam1991uncertainty, deffner2014optimal}. It consequently constrains the maximum speed of evolution that the system can reach. For a generic driven open quantum system, this parameter is defined as a unified expression based on the Schatten $p$ norm by Deffner and Lutz \cite{deffner2013quantum}:
\begin{equation}
\label{eq: QSLT}
    \begin{split}
 {\tau_\mathrm{QSLT}}=\max \left\{\frac{1}{\Lambda^\mathrm{op}_{\tau}}, \frac{1}{\Lambda^\mathrm{tr}_{\tau}}, \frac{1}{\Lambda^\mathrm{hs}_{\tau}}\right\} \sin^2(\mathcal{L}(\rho(0),\rho{(\tau)})),
    \end{split}
\end{equation}
$\mathcal{L}(\rho(0),\rho{(\tau)}) = \arccos \sqrt{\langle\psi_0|\rho{(\tau)}|\psi_0\rangle}$ is the so-called Bures angle between initial pure state $\rho(0)=\ket{\psi_0}\bra{\psi_0}$ and its evolved state $\rho{(\tau)}$, governed by the time-dependent master equation $\dot{\rho}(t)=L_t\rho(0)$ ($L_t$ is a super-operator). Therefore, we have $\sin^2(\arccos \sqrt{\langle\psi_0|\rho{(\tau)}|\psi_0\rangle} = 1-\langle\psi_0|\rho{(\tau)}|\psi_0\rangle$. It is worth mentioning that the Bures angle above is a measure of the distance between a pure state and a mixed state. The denominator in Eq.~(\ref{eq: QSLT}) indicates the average of $\dot{\rho}(t)$ over actual driving time duration $\tau$, i.e., $\Lambda^\mathrm{n}_{\tau}=\frac{1}{\tau}\int_{0}^{\tau}\mathrm{d}t \lVert \dot{\rho}(t) \rVert_\mathrm{n}$ (n $=$ $op$, $tr$, $hs$), with $op$, $tr$ and $hs$ denoting the operator, trace and Hilbert-Schmidt norms, respectively. It has been shown that the operator norm $\lVert \dot{\rho}(t) \rVert_\mathrm{op}={\displaystyle {\underset {i}{max}}}\{s_i\}$ always maximizes Eq.~(\ref{eq: QSLT}), where $s_i$ is the singular value of evolved density matrix $\dot{\rho}(t)$ \cite{jozsa1994fidelity, bures1969extension}. Thus, the QSLT in Eq.~(\ref{eq: QSLT}) can be simplified as
\begin{equation}
\label{b}
    \begin{split}
 \tau_\mathrm{QSLT}=\frac{1-\langle\psi_0|\rho{(\tau)}|\psi_0\rangle}{ \frac{1}{\tau}\int_{0}^{\tau}\mathrm{d}t~{\underset {i}{max}}\{s_i\}}.
    \end{split}
\end{equation}

The QSLT has the merit of evaluating the speed of quantum evolution for two regions: speedup and no speedup. The time-scaled notation of QSLT $(\tau_\mathrm{QSLT}/\tau)$ divides the evolution into two regions, speedup and no speedup respectively for $\tau_\mathrm{QSLT}/\tau<1$ and $\tau_\mathrm{QSLT}/\tau=1$. For the speedup region, it is revealed that as $\tau_\mathrm{QSLT}/\tau$ decreases, the process exhibits acceleration. However, the growth of $\tau_\mathrm{QSLT}/\tau$ results in the deceleration of the process. Regarding this significant result and with the aim of shedding light on the relationship between QSLT and non-Markovianity, we study the QSLT and non-Markovianity of a frequency-modulated qubit. Two scenarios have been considered depending on the initial state, first, the initial state is assumed to be an excited state, and second, it is a coherent superposition of ground and excited states.

\subsection{First scenario: Initial excited state} \label{subsec: Initial excited state}
We prepare the driven qubit initially in the excited state $\rho_1(0)= \ket {e}\bra {e}$ by taking parameters $\theta=\phi=0$ and substitute in Eq.~(\ref{b}), thus the QSLT reduces to
\begin{equation}
\label{c}
    \begin{split}
\frac{\tau_\mathrm{QSLT}}{\tau}=\frac{{1- {|{C}(\tau)|}^{2}}}{2\mathcal{N}{(\tau)}+{1- {|{C}(\tau)|}^{2}}},
    \end{split}
\end{equation}
where $\mathcal{N}{(\tau)}=\frac{1}{2}\bigg( \int_{0}^{\tau}|\partial_t {|{C}(t)|}^{2}| \mathrm{d}t + {|{C}(\tau)|}^{2} - 1 \bigg)$ is the BLP non-Markovianity measure and ${|{C}(\tau)|}^{2}$ characterizes the population of the excited state. As can be seen Eq.~(\ref{c}) indicates an analytical relationship between the QSLT and backflow of information. For the Markovian process $\mathcal{N}{(\tau)}=0$ and consequently $\tau_\mathrm{QSLT}/\tau=1$, which means the controlled quantum evolution has reached the highest possible speed and cannot speed the evolution further. However, for the non-Markovian case $\mathcal{N}{(\tau)}>0$ then $\tau_\mathrm{QSLT}/\tau<1$, which signifies the acceleration is possible so that, the smaller $\tau_\mathrm{QSLT}/\tau$, the larger potential capacity for speedup \cite{xu2018hierarchical}. 

\begin{figure}[t]
\begin{center}
    \includegraphics[width=0.48\textwidth]{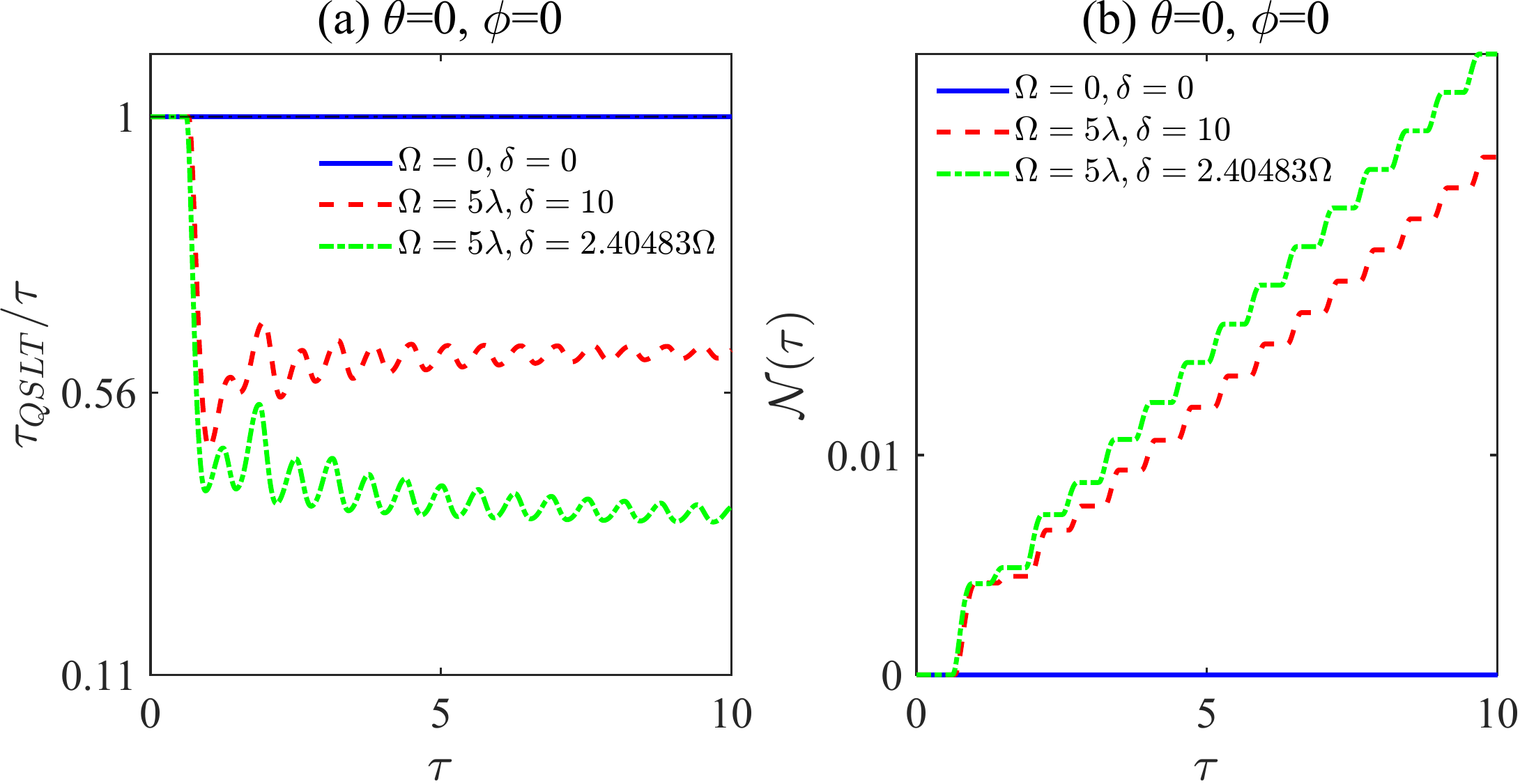}
\end{center} 
\caption{Ratio of the QSLT to the actual driving time $(\tau_\mathrm{QSLT}/\tau)$ and non-Markovianity $\mathcal{N}(\tau)$ as function of actual driving time $\tau$ for different modulations parameters: $\Omega=0$ and $\delta=0$ (blue solid line), $\Omega=5$ and $\delta=10$ (red dash line), and $\Omega=5$ and $\delta=2.40483\Omega$ (green dash-dot line). The excited state $\rho_1(0)=\ket {e} \bra {e}$ is considered as the initial state, and other parameters are $\gamma=0.1$ and $\lambda=1$.}
\label{QSL_tau}
\end{figure}

\begin{figure}[t]
\begin{center}
    \includegraphics[width=0.48\textwidth]{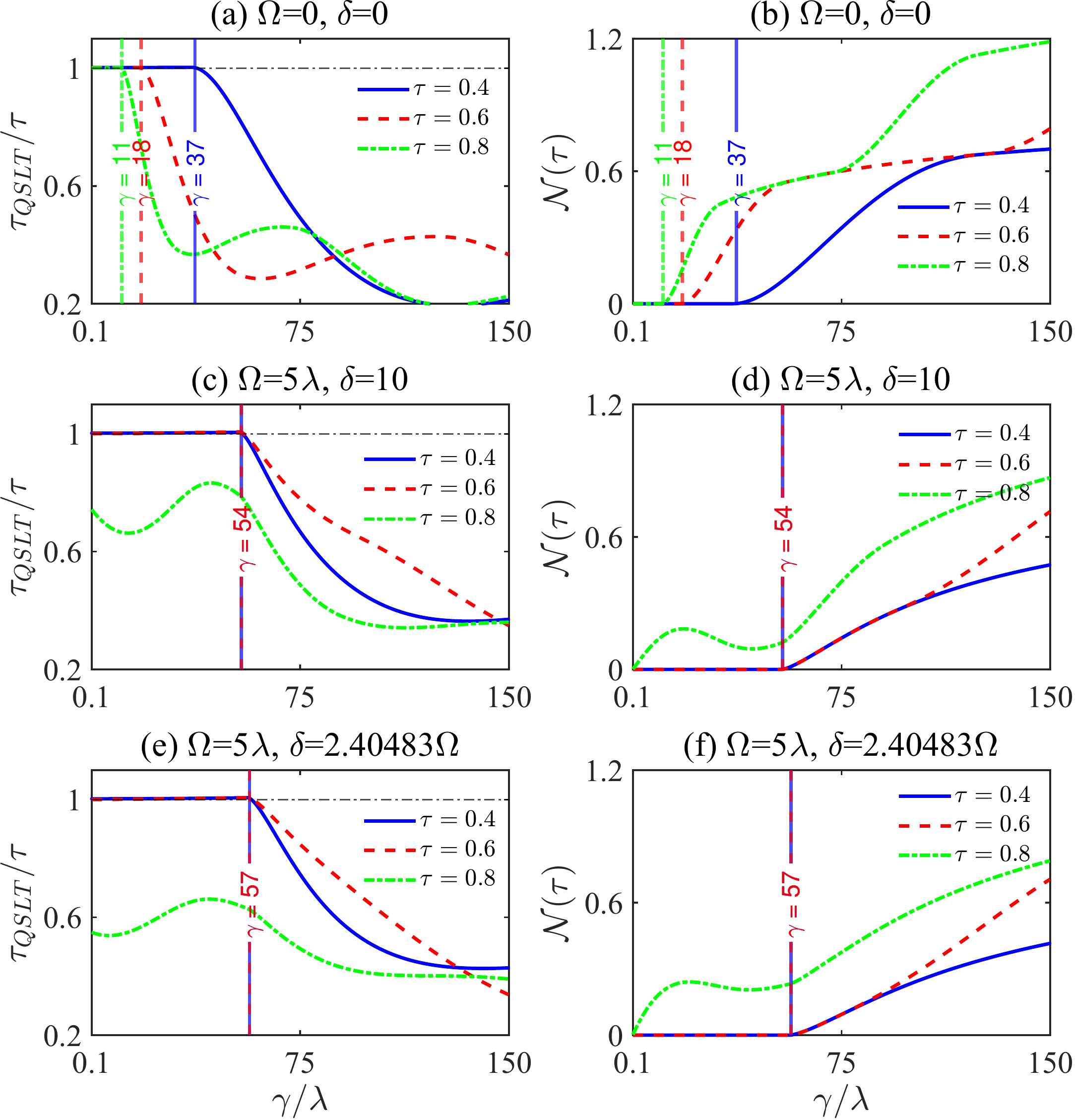}
\end{center} 
\caption{$\tau_\mathrm{QSLT}/\tau$ and non-Markovianity $\mathcal{N}(\tau)$ as a function of $\gamma/\lambda$ for different values of actual driving times $\tau=0.4$ (blue solid line), $\tau=0.6$ (red dash line), and $\tau=0.8$ (green dash-dot line) with $\lambda=1$. The excited state $\rho_1(0)=\ket {e} \bra {e}$ is considered as the initial state.}
\label{QSL_gamma}
\end{figure}

In Fig.~\ref{QSL_tau}, we plot the ratio of the QSLT to the actual driving time $(\tau_\mathrm{QSLT}/\tau)$ and non-Markovianity $\mathcal{N}(\tau)$ in terms of driving time for the coupling constant $\gamma=0.1$ and $\lambda=1$. It can be seen that the system initially remains in no speedup region and Markovian regime and then switches to speedup region and non-Markovian regime. It is noteworthy that when $\delta/\Omega$ is tuned to the zero of the Bessel function $J_0$, it not only appreciably accelerates the speedup process, but it increases the non-Markovianity as well. The results indicate that the frequency modulation parameters ($\delta$ and $\Omega$) have a key role in the speedup process.

\begin{figure}[t!]
\begin{center}
     \includegraphics[width=0.45\textwidth]{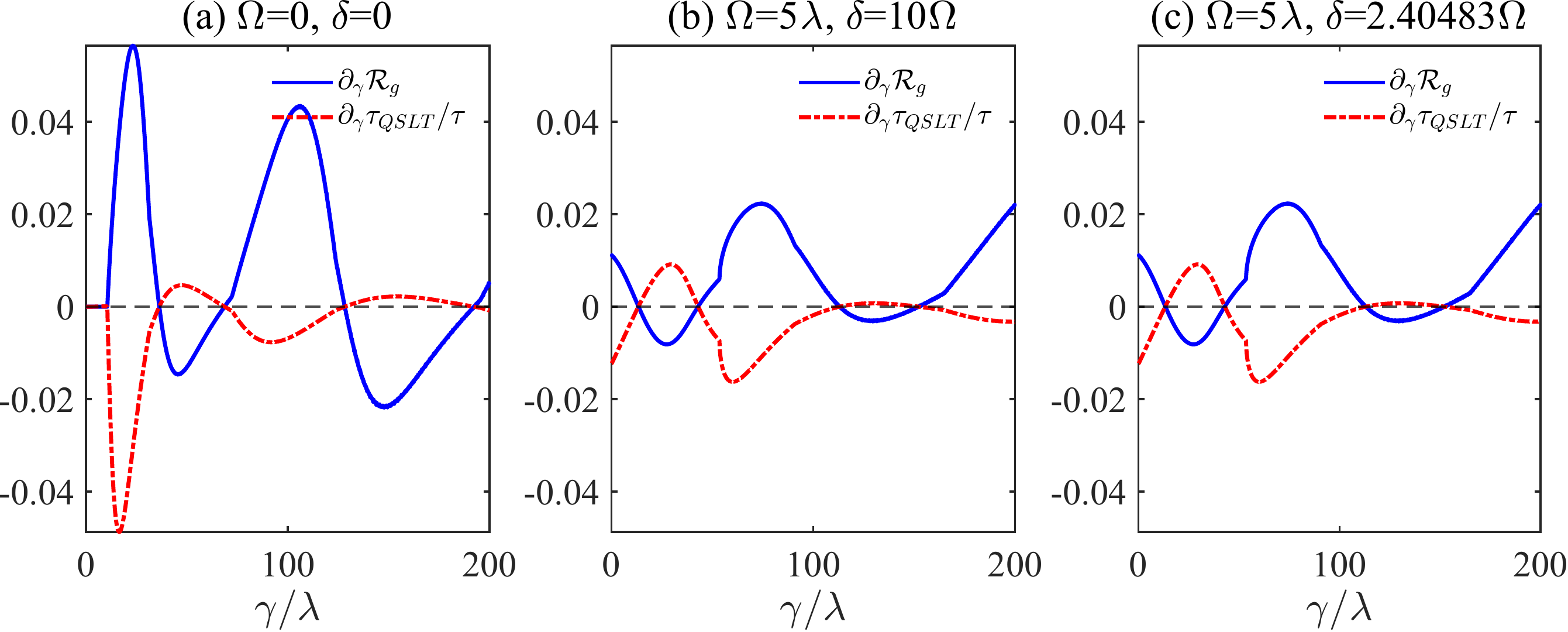}
\caption{Derivatives of $\tau_\mathrm{QSLT}/\tau$ (solid blue line) and $\mathfrak{R_g}$ (dash-dotted red line) as a function of $\gamma/\lambda$ for different actual driving time $\tau$ and $\lambda=1$. Dashed gray line is $y=0$. The excited state $\rho_1(0)=\ket {e} \bra {e}$ is considered as the initial state.}
\label{AR7}
\end{center}
\end{figure}

To gain more detailed information concerning the link between QSLT and non-Markovianity, we plot  $\tau_\mathrm{QSLT}/\tau$ and non-Markovianity as a function of $\gamma/\lambda$ for different values of driving time $\tau$ under different modulation frequencies $\Omega$ and amplitudes $\delta$ in Fig.~\ref{QSL_gamma}. As it is evidently displayed, there are specific values of the $\gamma/\lambda$ for which the system makes the transition from no speedup to speedup regions. Moreover, it is surprisingly disclosed that the switching point from Markovian to non-Markovian regimes perfectly matches the transition point, albeit such switching would occur regardless of the initial state. Since the excited state population appears in the  Eq.~(\ref{c}), we wonder how this parameter affects the QSLT. It is speculated that the QSLT is affected by an interplay between non-Markovianity and the population of the excited state during the evolution. Hence, we introduce a new parameter $\mathfrak{R_g}$ which illustrates this interplay as
\begin{equation}
 \mathfrak{R_g}:=\frac{\mathcal{N}(\tau)}{1-\langle\psi_0|\rho_{\tau}|\psi_0\rangle}=\frac{\mathcal{N}(\tau)}{1-|{C}(\tau)|^{2}}.
\end{equation}
As this ratio suggests, it is far better to discuss the collective effect of the non-Markovianity and population of the excited state on the QSLT to find an explicit relationship. In order to support this claim, we plot derivatives of $ \mathfrak{R_g}$ and $\tau_\mathrm{QSLT}/\tau$ as a function of $\gamma/\lambda$ for different values of actual driving time $\tau$ under different values of modulation amplitudes $\delta$ and frequencies $\Omega$ in Fig.~\ref{AR7}. As the curves illustrate, there is approximately mirror symmetry between the derivations of $\mathfrak{R_g}$ and $\tau_\mathrm{QSLT}/\tau$ conveying this message that an increase in the ratio $\mathfrak{R_g}$ can always lead to the speedup in the quantum evolution.

\subsection{Second scenario: Initial coherent superposition state} 

\begin{figure}[t]
\begin{center}
    \includegraphics[width=0.48\textwidth]{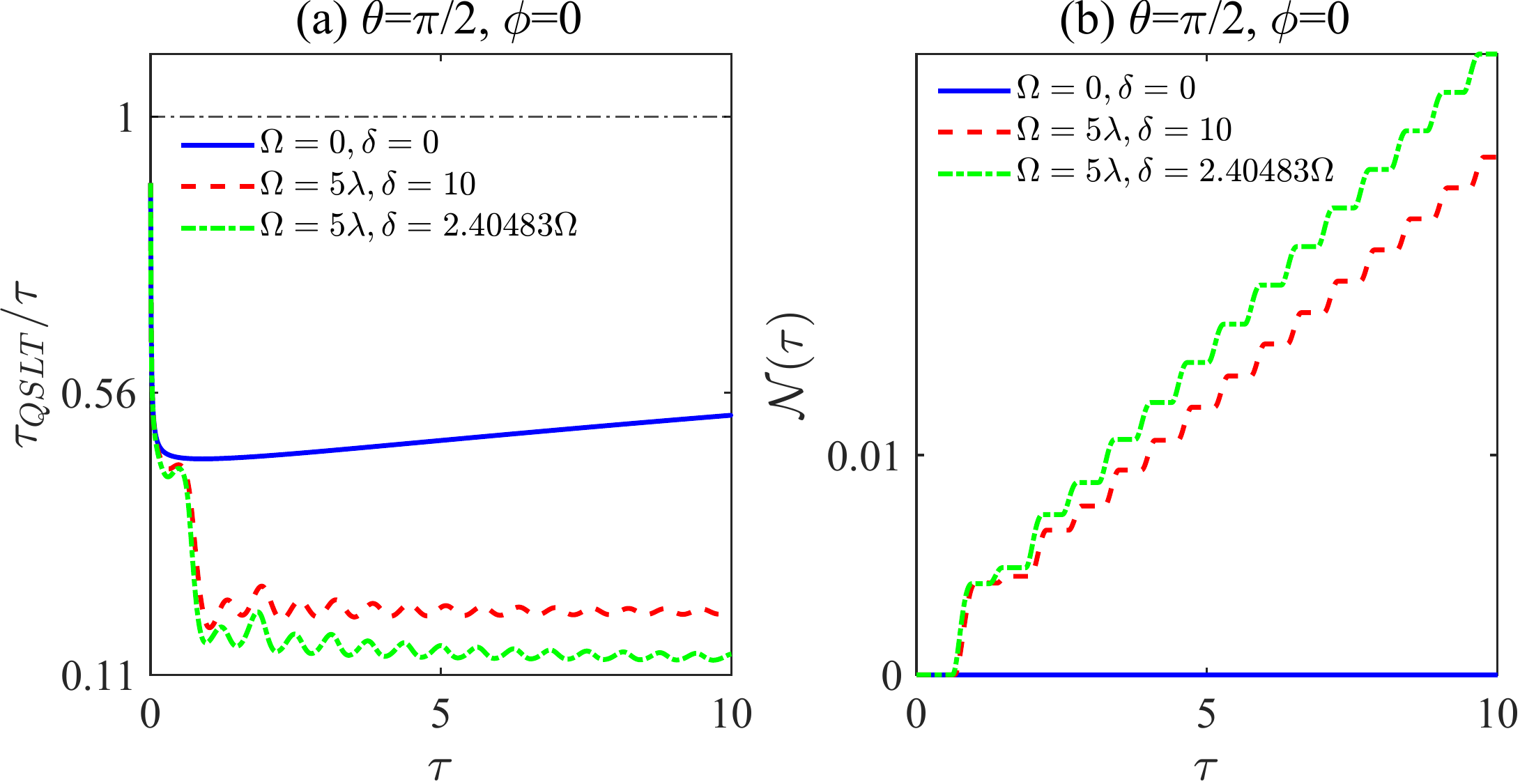}
\end{center} 
\caption{$\tau_\mathrm{QSLT}/\tau$ and non-Markovianity $\mathcal{N}(\tau)$ as a function driving time $\tau$ for different modulations parameters: $\Omega=0$ and $\delta=0$ (blue solid line), $\Omega=5$ and $\delta=10$ (red dash line), and $\Omega=5$ and $\delta=2.40483\Omega$ (green dash-dot line). The excited state $\rho_1(0)=\ket {e} \bra {e}$ is considered as the initial state, and other parameters are $\gamma=1$ and $\lambda=3$.}
\label{QSL_tau_2}
\end{figure}

\begin{figure}[t!]
\begin{center}
    \includegraphics[width=0.48\textwidth]{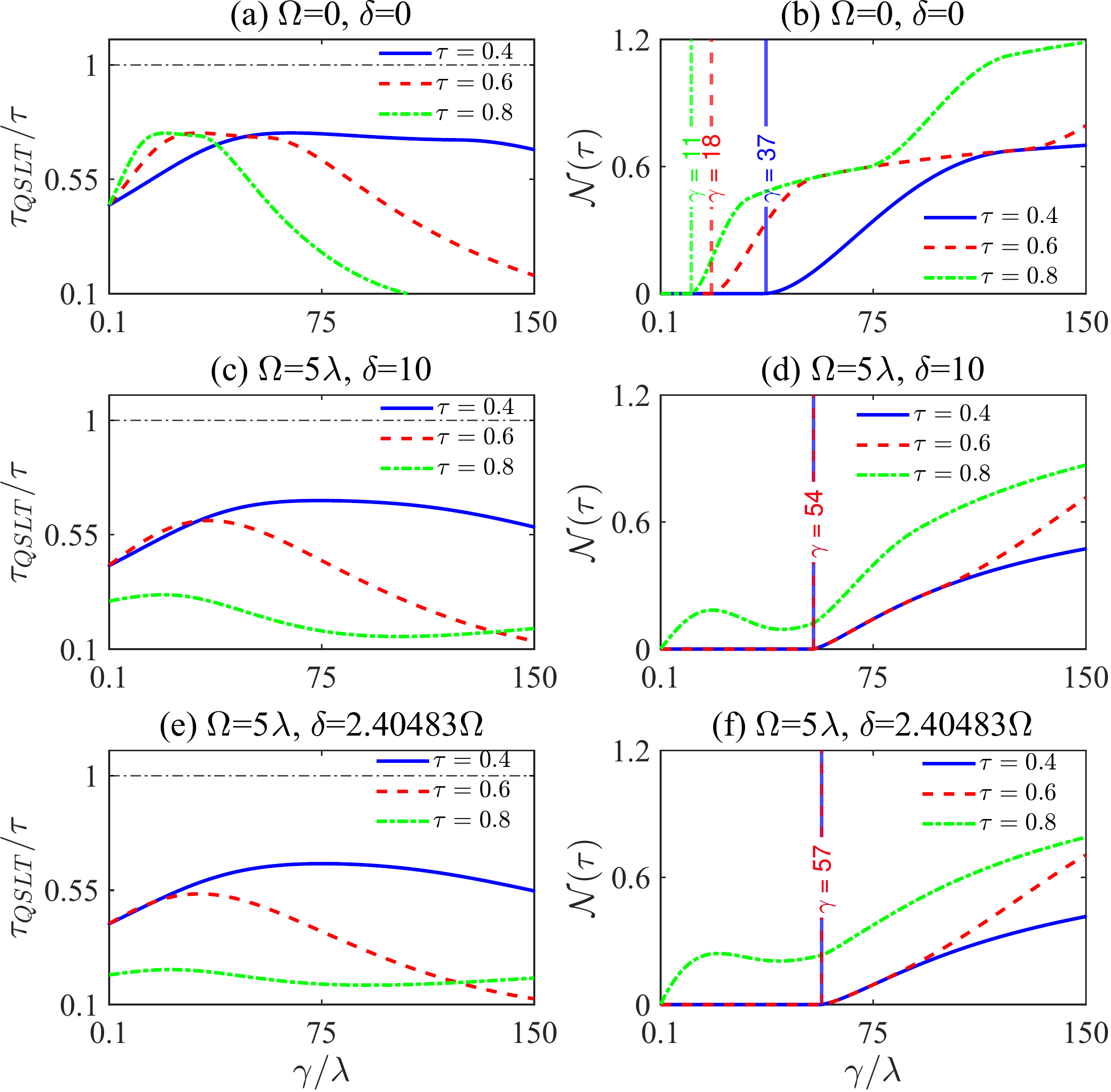}
\end{center} 
\caption{$\tau_\mathrm{QSLT}/\tau$ and non-Markovianity $\mathcal{N}(\tau)$ as a function of $\gamma/\lambda$ for different values of actual driving time $\tau=0.4$ (blue solid line), $\tau=0.6$ (red dash line), and $\tau=0.8$ (green dash-dot line) with $\lambda=1$. The coherent superposition $\ket{\psi(0)} = \frac{1}{\sqrt{2}}\left(\ket{e}+\ket{g}\right)$ is considered as the initial state.}
\label{AR10}
\end{figure}

\begin{figure}[t]
\begin{center}
    {\includegraphics[width=0.48\textwidth]{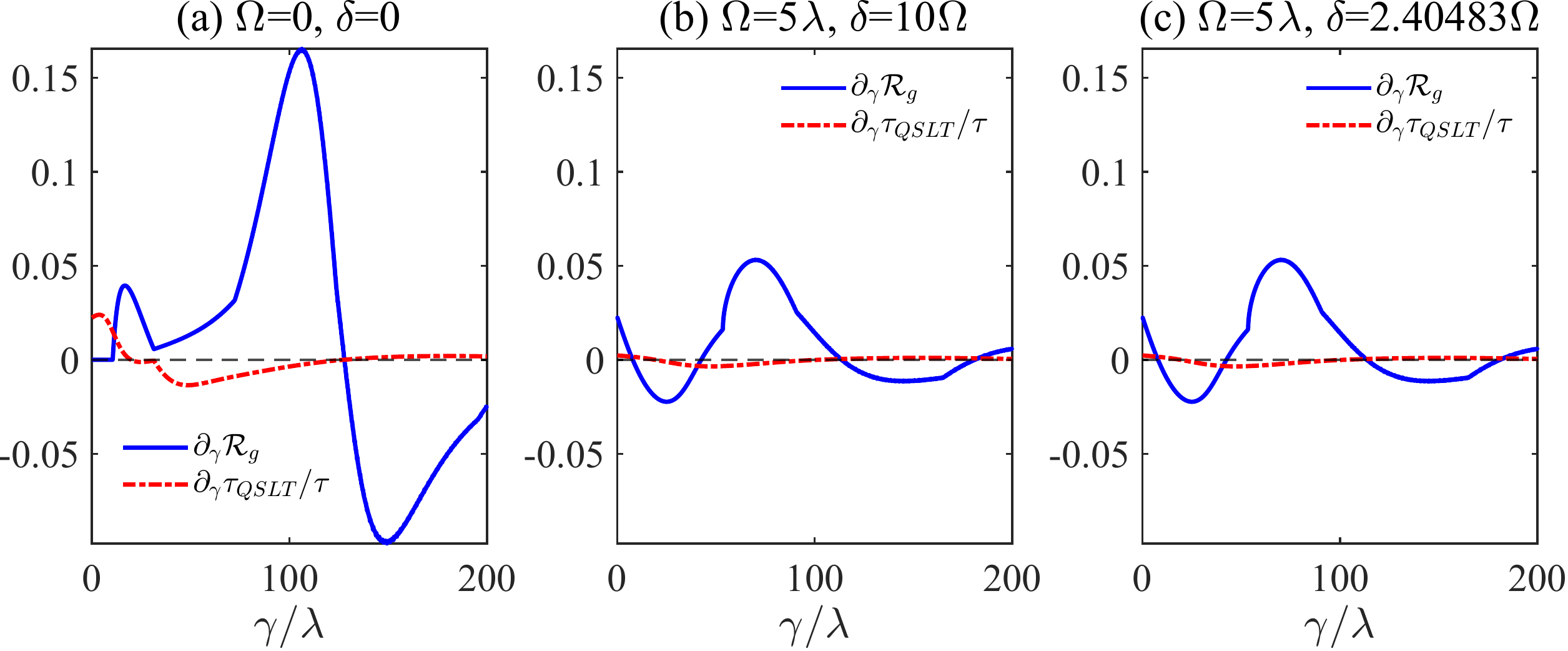}}
\end{center} 
\caption{Derivatives of $\tau_\mathrm{QSLT}/\tau$ (solid blue line) and $\mathfrak{R_g}$ (dash-dotted red line) as a function of $\gamma/\lambda$ for actual driving time $\tau = 1$ with $\delta=5$, $\Omega = 5$, and $\lambda=1$. Solid gray line is $y=0$. The superposition state $\rho_2(0) = \frac{1}{2}(\ket{e}\bra{e}+\ket{e}\bra{g}+\ket{g}\bra{e}+\ket{g}\bra{g})$ is considered as the initial state.}
\label{AR13}
\end{figure}

We resume the process considering the second scenario according to which the system starts from a coherent superposition $\ket{\psi(0)} = \frac{1}{\sqrt{2}}\left(\ket{e}+\ket{g}\right)$ by taking parameters $\theta=\pi/2, \phi=0$. To carry out a comparative study, we proceed with the parallel trend with the previous subsection. Thus, from Eq.~(\ref{b}), the QSLT is given by
\begin{equation}
\label{d}
    \begin{split}
 \frac{\tau_{QSL}}{\tau}=\frac{1-\mathrm{Re}(C({\tau}))}{\int_{0}^{\tau}\mathrm{d}t\left|\sqrt{\left|\partial_t C(t)\right|^2+|\partial_t |C(t)|^2|^2}\right| }.
    \end{split}
\end{equation}
As is clear, no explicit expression is found for $\tau_\mathrm{QSLT}/\tau$ in terms of non-Markovianity. The dynamic behavior of $\tau_\mathrm{QSLT}/\tau$ and non-Markovianity versus actual driving time is respectively depicted in Fig.~\ref{QSL_tau_2} (a) and (b). It is revealed that, by initially preparing the system in a coherent superposition state, the evolution of system is already laid in the speedup region whether the qubit is frequency modulated or not. Nonetheless, for the case qubit is unmodulated, $\tau_\mathrm{QSLT}/\tau$ plummets down initially and then gradually increases which indicates deceleration. However, for the frequency-modulated qubit, $\tau_\mathrm{QSLT}/\tau$ at first sharply decreases, and after a short while with the commencement of non-Markovian behavior, falls one more step and then incessantly fluctuates implying constant acceleration and deceleration. Fig.~\ref{AR10} exhibits the behavior of $\tau_\mathrm{QSLT}/\tau$ and non-Markovianity $\mathcal{N}(\tau)$ as a function of $\gamma/\lambda$ for different values of actual driving time $\tau$ under different modulation frequencies $\Omega$ and amplitudes $\delta$. As it is manifest, the dynamic of the system for all values of $\gamma/\lambda$ is laid in the speedup region regardless of the modulation process. However, modulation exerts an influence on the non-Markovian behavior. Nevertheless, one can notice that as the system switches from the Markovian to the non-Markovian regimes, the curves of $\tau_\mathrm{QSLT}/\tau$ experience a declining trend signifying acceleration. Moreover, as it is evident, the modulation process gives rise to a marked shift in the initial value of $\tau_\mathrm{QSLT}/\tau$ for $\tau=8$ the same as the case where the system is initially prepared in the excited state. As a final investigation, we plot the derivatives of $\mathfrak{R_g}$ and $\tau_\mathrm{QSLT}/\tau$ versus $\gamma/\lambda$ in Fig.~\ref{AR13}. As it is crystal clear, when the system is initially prepared in the coherent superposition state, there is no significant relationship between $\mathfrak{R_g}$ and speedup process.

\section{Conclusion} \label{secfive}

In this work, we have carried out a thorough study of the quantumness and speedup limit time of a frequency-modulated qubit embedded in a leaky cavity. The validity of the standard quantum witness as a quantum indicator is checked for the proposed system in comparison to the certain criterion of quantumness namely, coherence monotone.  The proposed model best exemplifies that SQW fails to truly describe the non-classicality of a generic non-isolated system (including non-Markovian environmental effect) certifying the result of ref \cite{21.Farzam2020}.  Moreover, the results indicate that, in contrast to the previous reports in which SQW could not reach the coherence monotone, the SQW can exceed the coherence monotone as well. Such glaring inconsistency between the behaviors of SQW and coherence monotone leads us to employ the optimization method, introduced in Ref.~\cite{21.Farzam2020}, by choosing the appropriate blind measurement and proceeding according to the optimal quantum witness. We have demonstrated that OQW keeps up with the coherence monotone and its maximum values perfectly coincide with the coherence monotone for the different values of modulation parameters.

We have then assessed the impact of the frequency modulation on the speedup evolution and shed light on the relationship between QSLT and non-Markovianity, by considering two cases depending on the initial state. Surprisingly, it has emerged that the non-Markovianity is not the sole influential factor in speeding up the rate of evolution but the population of the excited state also participates as a contributing factor. Hence, we have introduced a new parameter $\mathfrak{R_g}$, defined as the ratio of non-Markovianity to the population of the excited state, for exploring its relationship with the usual ratio of QSLT to the actual driving time.
The findings indicate that the interplay of modulation parameters $\delta$ and $\Omega$ has a key role in the speedup process.

For the case when the initial state is the excited state, we have observed that the transition point from no speedup to speedup regions perfectly matches to switching point from Markovian to non-Markovian regimes. We have also found that the relationship between $\mathfrak{R_g}$ and $\tau_\mathrm{QSLT}/\tau$ conforms to a meaningful pattern so that as the $\mathfrak{R_g}$ increases, the ratio of $\tau_\mathrm{QSLT}/\tau$ decreases implying the acceleration of speedup process. However, when the initial state is assumed to be a coherent superposition of the ground and excited states, the evolution of system is already laid in the speedup region whether the process is Markovian or non-Markovian and there is no significant relationship between $\mathfrak{R_g}$ and $\tau_\mathrm{QSLT}/\tau$. Therefore, all together the achieved results verify our intuition about the effect of the excited state population on the QSLT. Also, as a common result for both cases regardless of the initial state, non-Markovianity induces fluctuation of the speedup process implying periodic acceleration and deceleration. 

As a prospect, the optimal quantum witness can be potentially applied to the complex quantum systems such as the flux qubit \cite{knee2016strict} and IBM quantum computer \cite{ku2020experimental}, for which the implementation of quantum tomographic methods is problematic to indicate quantum coherence. Moreover, frequency modulation can be employed as a simple but effective technique to accelerate the process of performing quantum tasks. As a matter of fact, experimentalists have recently utilized this technique to fabricate and control quantum circuit devices in superconducting Josephson qubits (artificial atoms) \cite{56.Silveri2017, 57.Nakamura2001, oliver2005mach, 59.Tuorila2010, 60.Tuorila2013}, enabling progress in the building blocks of current quantum computer prototypes \cite{61.Quantumcomputing}. Hence, qubits with optimal transition frequency modulation are good candidates for quantum gates owing to the fact that they can perform fast quantum operations.

\begin{acknowledgments}
A.M. acknowledges the support of the University of Guilan. R.M. acknowledges support from NSERC, MEI and the CRC program in Canada. R.L.F. acknowledges support from Unione europea -- NextGenerationEU -- fondi MUR D.M. 737/2021 -- progetto di ricerca ``Indistinguibilit\`{a} come risorsa controllabile per processi di informazione quantistica e termodinamica quantistica''.
\end{acknowledgments}


\end{document}